\documentclass[aps,showpacs,twocolumn]{revtex4}
\usepackage{color,graphicx,subfig}
\usepackage{epsfig,amsmath}
\usepackage{xcolor}
 
\begin{document}
\title{Strangeness-driven phase transition in (proto)-neutron star matter}

\author{F. Gulminelli$^{1}$ }
\author{Ad. R. Raduta$^{2}$}
\author{M. Oertel$^{3}$} 
\author{J. Margueron$^{4}$}

\affiliation{$^{1}$~CNRS and ENSICAEN, UMR6534, LPC,
14050 Caen c\'edex, France\\
$^{2}$~IFIN-HH, Bucharest-Magurele, POB-MG6, Romania\\
$^{3}$~LUTH, CNRS, Observatoire de Paris, 
Universit\'e Paris Diderot, 5 place
Jules Janssen, 92195 Meudon, France\\
$^{4}$~Institut de Physique Nucl\'eaire, IN2P3-CNRS, 
Universit\'e  Paris-Sud, F-91406 Orsay cedex, France.
}

\begin{abstract}
  The phase diagram of a system constituted of neutrons, protons,
  $\Lambda$-hyperons and electrons is evaluated in the mean-field
  approximation in the complete three-dimensional space given by the baryon,
  lepton and strange charge.  It is shown that the phase diagram at
  sub-saturation densities is strongly affected by the electromagnetic
  interaction, while it is almost independent of the electric charge at
  supra-saturation density. As a consequence, stellar matter under the
  condition of strangeness equilibrium is expected to experience a first as 
  well as a second-order
  strangeness-driven phase transition at high density, while the liquid-gas
  phase transition is expected to be quenched. An RPA calculation indicates
  that the presence of this critical point might have sizable implications
  for the neutrino propagation in core-collapse supernovae.
\end{abstract}


\today

\maketitle

\section{Introduction}

Supernova explosions following the gravitational collapse of a massive
star ($M \gtrsim 8 M_\odot$) are among the most fascinating events in
the universe as they radiate as much energy as the sun is expected to
emit over its whole life span \cite{bethe}. Nuclear
physics is an essential ingredient in the numerical simulations which
aim to describe these events, since realistic astrophysical
descriptions of the collapse and post-bounce evolution 
rely on the accuracy
of the implementation of weak processes 
 and equation(s) of state (EOS) \cite{Janka2006,prakash}.  
Determining the EOS constitutes a particularly difficult task
since phenomenology ranges from a quasi-ideal un-homogeneous gas to
strongly interacting uniform matter and, potentially, deconfined
quark-matter.  The situation is even more difficult if phase
transitions are experienced, since mean-field models fail in such
situations \cite{glendenning}.

The Coulomb-quenched liquid-gas (LG) phase transition taking place at
densities smaller than the nuclear saturation density 
($n_0 = 0.16$ fm$^{-3}$) is, probably, the most notorious and best
understood case \cite{Ducoin-06,barranco,shlomo,haensel,rios}. 
 At highest densities, a quark-gluon plasma is expected, 
but predictions on the exact location of the transition  are strongly 
model dependent \cite{Sagert}. 
 In the intermediate density domain 
 simple energetic considerations show that additional degrees
 of freedom may be available, such as hyperons, nuclear resonances,
 mesons or muons \cite{glen82}. The possibility that the onset of
 hyperons could pass via a first order phase transition in neutron
 stars has been evoked in Ref. \cite{SchaffnerBielich2002}, using a
 relativistic mean field model (RMF), and in
 Ref. \cite{SchaffnerBielich2000}, a phase transition between phases
 with different hyperonic species has been observed for cold
 matter. 
 The possibility of a first order phase transition to hyperonic
   matter in effective RMF models has been discussed in 
   Refs.~\cite{Wang2004,Yang2003,Yang2005},
   too. Within the latter models, the phase transition region is 
   located at sub-saturation densities, and is thus not relevant for star
   matter. 
Using a simple two-component $(n,\Lambda)$ model, we have
 recently studied the complete phase diagram of strange baryonic
 matter showing that it exhibits a complex structure with first and
 second order phase transitions \cite{nL}.  However, the exploratory
 calculation of Ref. \cite{nL} neglects the
 fact that in addition to baryon number $B$ and strangeness $S$, 
 the charge $Q$ and lepton $L$ quantum numbers are also populated.  
The thermodynamics of the complete system should thus be 
studied in the four-dimensional space of the associated 
charges $n_B,n_S,n_L,n_Q$. The strict electroneutrality constraint
$n_Q=0$, necessary to obtain a thermodynamic limit \cite{anomalous},
makes the physical space three-dimensional. As it is known from
the EOS studies at sub-saturation density \cite{Ducoin-07}, 
the introduction of
the charge degree of freedom can have a very strong
influence on the phase diagram and cannot be neglected.
In this work we therefore introduce a four-component model
constituted of neutrons, protons, electrons and $\Lambda$-hyperons.
Electrons are treated as an ideal gas.

We present, in sec.~\ref{sec:thermo} of this paper, the thermodynamics and phase
transition of the n, p, e and $\Lambda$ system, and discuss the
influence of the Coulomb interaction. The consequence of the
phase transition on the cooling of proto-neutron stars, through the
neutrino mean free path, is qualitatively discussed in sec.~\ref{sec:neutrinos}.
Finally, we present our conclusions in sec.~\ref{sec:conclusions}.

\section{Thermodynamics of a n, p, $\Lambda$ system with electrons}
\label{sec:thermo}
In the widely used mean-field approximation
\cite{glen82,stone,oertel2012,weissenborn11,bednarek11,hofmann2001,bonanno11,
grigorian} 
the total baryonic energy density  is given by the sum of the 
mass, kinetic and potential  energy density functionals which represents 
a surface in the three-dimensional space defined by the baryon, 
strange and charge density 
given, in our case, by
$n_B=n_n+n_p+n_{\Lambda}$,
$n_S=- n_{\Lambda}$ and
$n_Q=n_p$. In the
non-relativistic formalism valid in the considered domains of density
and temperature it reads
\begin{equation}
 e_B=\sum_{i=n,p,\Lambda}\left ( n_i m_i c^2+ \frac{\hbar^2}{2m_i} \tau_i \right )
+ e_{pot}(n_n,n_p,n_{\Lambda})
\end{equation}

The single-particle densities are given by the Fermi
integrals
\begin{equation}
n_i
=\frac{4 \pi }{h^3} \left(\frac{2m_i}{\beta} \right)^{\frac 32} 
F_{\frac 12}(\beta \tilde \mu_i)  ; \;
\tau_i=\frac{8\pi^3 }{h^5} \left( \frac{2 m_i}{\beta } \right )^{\frac 52}
F_{\frac 32}(\beta \tilde \mu_i),
\end{equation}
where $F_{\nu}(\eta)=\int_0^{\infty} dx \frac{x^{\nu}}{1+\exp \left(
    x-\eta\right)}$ is the Fermi-Dirac integral, $\beta=T^{-1}$ is
the inverse temperature, $m_i$ is the
effective $i$-particle mass and $\tilde \mu_i$ is the effective chemical
potential of the $i$-species self-defined by the single-particle density.

\subsection{The model}

A full thermodynamics characterization of the system
is provided by the  pressure 
$P_B = Ts_B - e_{B} + \sum_i \mu_i n_i$ 
together with the entropy density  $s_B$ in mean-field,
\begin{equation}
s_B=\sum_{i,p,\Lambda}
\left[ \frac{10 \hbar^2}{6\, m_i} \beta \tau_i
-n_i \beta \tilde \mu_i\right]~.
\label{eq:entropy}
\end{equation}

The thermodynamical definition  $n_i \doteq
\left(\frac{\partial P}{\partial \mu_i}\right) |_{\beta}$ allows to infer the
relation among the  chemical potentials $\mu_i$ and the effective
parameters $\tilde \mu_i$ as $\mu_i=\tilde \mu_i + m_i c^2 +U_i$, with
$U_i=\partial e_{pot}/\partial n_i$.

Within the numerical applications we shall use the potential energy
density proposed by Balberg and Gal~\cite{BG},

\begin{eqnarray}
e_{\mathit{pot}}(n_n,n_p,n_\Lambda) &=& \sum_{i,j=\{n,p,\Lambda\}} \left(a_{ij}
  n_i n_j + b_{ij} t_i t_j n_i n_j \right. \\ && \left.+ c_{ij} 
\frac{1}{n_i + n_j} (
  n_i^{\gamma_{ij}+1} n_j + n_j^{\gamma_{ij}+1} n_i) \right)~,\nonumber 
\label{eq:epspot}
\end{eqnarray} 
accounting for nucleon-nucleon, nucleon-$\Lambda$ and $\Lambda$-$\Lambda$
interactions.  $t_i$ denotes the third isospin component of particle
$i$. In the non-strange sector the form of the interaction is the
same as in the widely used Lattimer-Swesty \cite{LS} EOS. 

Let us mention that the observation of a neutron star (PSR J 1614-2230)
  with a mass of almost two solar masses~\cite{Demorest2010} imposes stringent
  constraints on the hyperonic interaction in dense neutron star matter. The
  maximum mass for a $n,p,\Lambda+e$ system as studied in the present
  manuscript is $2.04 M_\odot$ with the parameter set BG I for the coupling
  constants (see Table~\ref{table:bg1}) in agreement with the mass of PSR J
  1614-2230.  Including all hyperonic degrees of freedom, the maximum neutron
  star mass obtained with parametrisation BG I decreases and becomes slightly
  too low. However, the qualitative results discussed here about the
  thermodynamics of the system and the consequences on the neutrino mean free
  path are independent of the parametrisation used. In particular, the same
  qualitative results are obtained with the parametrisations from
  Ref.~\cite{oertel2012}, in agreement with the mass of PSR J 1614-2230 even
  upon including all the different hyperons. Quantitative differences are very
  small, such that we have chosen here to use for numerical applications one
  parametrisation from the original paper by Balberg and
  Gal~\cite{BG}, BG I.

\begin{table*}
\begin{center}
\caption{\label{table:bg1}Coupling constants corresponding to the stiffest interaction 
proposed in Ref.~\cite{BG}.}
\begin{tabular}{cccccccccc}
\hline
\hline
Parameter set & $a_{NN}$ & $b_{NN}$  & $c_{NN}$ &
$a_{\Lambda\Lambda}$ & $c_{\Lambda\Lambda}$ & $a_{\Lambda N}$ &
$c_{\Lambda N}$ & $\gamma_{NN}$ & $\gamma_{\Lambda N}$ \\
 & MeV fm$^3$ & MeV fm$^3$   & MeV fm$^{3\delta}$  &  MeV fm$^3$  &
MeV fm$^{3\gamma_{NN}}$  & MeV fm$^3$  &  MeV fm$^{3\gamma_{\Lambda N}}$  &  & \\
\hline
BGI             &-784.4 &  214.2 & 1936. & -486.2 & 1553.6 & -340. &
1087.5 & 2 & 2 \\
\hline
\hline
\end{tabular}
\end{center}
\end{table*}

\subsection{Instabilities and phase-transition}

Matter stability with respect to phase separation 
can be checked in any point of the extensive variable
space by analyzing the eigen-values of the curvature matrix 
\cite{Margueron2003b,Ducoin-06,Ducoin2008},
$C_{ij}=
\partial^2 f(\{n_l\}_{l=\{i,j,k\}})/\partial n_i \partial n_j 
$, where $i,j,k=B,S,Q$ and $f=e_{tot}-T s_{tot}$ 
is the total free-energy.
The occurrence of, at least, one negative eigen-value in a certain domain of
$(n_B,n_S,n_Q)$ means that the system is unstable versus phase
separation. 
The associated  $3$-dimensional Gibbs construction can be reduced to a simpler
1-dimensional Maxwell construction \cite{Ducoin-06} by performing a Legendre
transformation with respect to two out of the three chemical potentials 
$\mu_B=\mu_n$, 
$\mu_S=- \mu_{\Lambda}+\mu_n$ and
$\mu_Q=\mu_p-\mu_n$.
We have chosen to work in the hybrid ensemble $(n_B,\mu_S,\mu_Q)$ defined by:
\begin{equation}
\bar f_{baryon}(n_B, \mu_S, \mu_Q)=
f_{baryon}-\mu_S n_S - \mu_{Q} n_{Q}, \label{hybrid}
\end{equation}

\begin{figure}[ht]
\begin{center} 
\includegraphics[width=0.95\columnwidth]{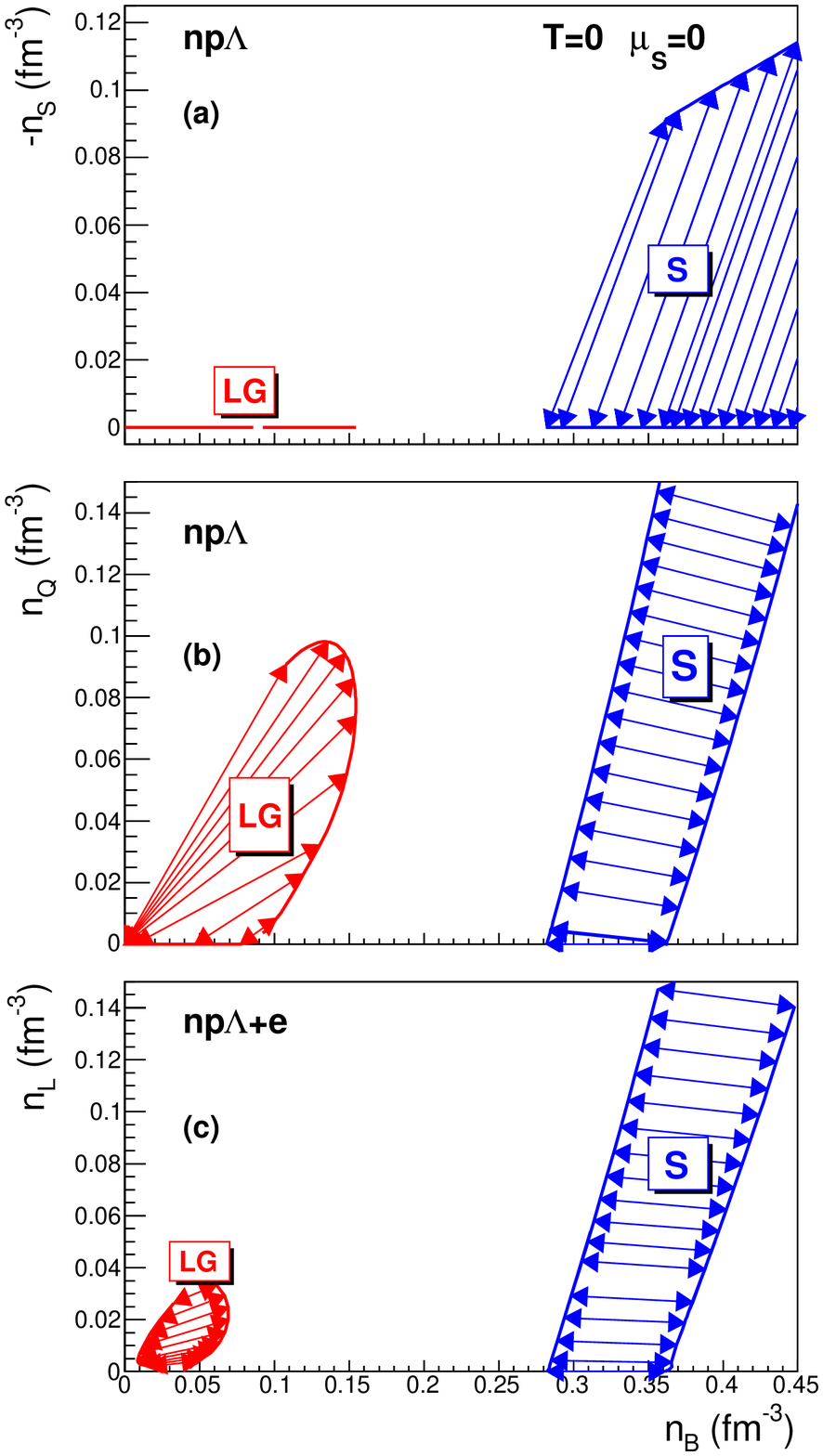}
\end{center}
\caption{\small (Color online) 
Borders of the phase-coexistence domains at T=0 and $\mu_S=0$.
Upper (middle):  $(n, p, \Lambda)$-mixture in
$n_B-n_S$ ($n_B-n_C$) coordinates.
Lower:  $(n, p, \Lambda,e)$-mixture in $n_B-n_L$ coordinates.
Red: liquid-gas phase transition of non-strange dilute nuclear matter;
blue: non-strange to strange  phase transition.
The arrows mark the directions of phase separation.}
\label{Fig1}
\end{figure} 

If the associated equation of state $\mu_B=\partial \bar
  f_{baryon}(n_B, \mu_S, \mu_Q)/\partial n_B$ as a function of $n_B$
  presents a slope inversion, the relation $\mu_B(n_B)$ is
  three-valued within a given interval of $\mu_B$. Then a Maxwell
  equal-area construction on this function allows defining two values
  $n_B^{(1)}$, $n_B^{(2)}$, which are characterized by the complete
  Gibbs equilibrium conditions $\left
  (P,\mu_B,\mu_S,\mu_Q\right)_{(1)}=\left
  (P,\mu_B,\mu_S,\mu_Q\right)_{(2)}$ for a multi-component system
  . The hyper-surface connecting the two points $\left
  (n_B,n_S,n_Q\right)_{(1)}$,$\left (n_B,n_S,n_Q\right)_{(2)}$ is the
  usual Gibbs construction. This procedure is independent of the
  choice of the densities (here: $n_S,n_Q$) to be
  Legendre-transformed, provided the order parameter has a
  non-vanishing component along the remaining density (here:
  $n_B$). If this was not the case, that is if there was no jump in
  $n_B$ at the phase transition, $n_B^{(1)}=n_B^{(2)}$, the
  information on the phase transition could not be extracted from the
  hybrid ensemble eq.(\ref{hybrid}). We have verified that this is
  never the case, and the phase transition we will identify always
  separates a more diluted (lower $n_B$) from a denser (higher $n_B$)
  phase. For a generic physical multi-component system, this is not
  always the case and in the general case the convexity properties of
  the free energy have to be examined with care in order to identify
  phase transitions in such systems. In particular for our specific
  physics application of strangeness phase transition, we will
  explicitly show that the charge density is almost unaffected by the
  phase transition. This means that the concavity of the free energy
  $\bar f_{baryon}(\mu_B, \mu_S, n_Q)$ as a function of $n_Q$ is
  extremely small. This means that working in that statistical
  ensemble would have rendered the observation of the phase transition
  very difficult.  We stress that the one-dimensional Maxwell
  construction in the hybrid ensemble eq. (\ref{hybrid}), as long as a
  $n_B$ jump occurs through the phase transition as it does here, is
  strictly equivalent to the complete Gibbs construction.  In
  particular, the pressure, as function of one density with the other
  densities kept constant, has not a constant value in the mixed
  phase~\cite{Ducoin-06}. For instance, $P(n_B)$ at fixed $n_S, n_Q$
  is not constant in the mixed phase region.  On the contrary, a
  Maxwell construction on $P(n_B)$ or $\mu_B(n_B)$ at constant values
  of $n_S,n_Q$ is never theoretically justified.  

The upper part of Fig. \ref{Fig1} illustrates the projection of the
$T=0$ phase diagram in the $n_B-n_S$ plane for $\mu_S=0$. 
{The arrows mark the direction of phase separation which, in case of
phase coexistence, coincides with the order parameter.}
Two phase-coexistence domains may be identified.  
The one lying along $n_S=0$ at
sub-saturation density corresponds to the well known LG like phase
transition taking place in dilute nuclear matter \cite{Ducoin-06}.  
The second domain lies at supra-saturation densities ($n_B\gtrsim 2 n_0$), 
and the direction of phase separation is dominated by the strange
density. 
In the density range shown by the figure, this domain is  not upper limited in $\rho_B$. 
We observe a bending at very high density meaning that the domain is
  finite as for the sub-saturation LG transition, but since we do not consider
  these densities as being realistically described within the present model,
  we refrain from showing the whole domain here.
This transition is consistent with our previous findings within a
simpler 2D model \cite{nL}, 
though it obviously depends on the assumed strengths of the $NN, NY$ and 
$YY$-interactions. 
It is at first sight surprising to observe that the coexistence border 
of the latter transition is given by simple straight lines.  In principle the 
coexistence borders of a first order phase transition with three conserved
charges are given by two surfaces in the three dimensional space. For their
projection on the $n_B-n_S$ plane to be given by a one-dimensional curve,
these surfaces have to be perpendicular to that plane,
that is independent of $n_Q$.
The observed independence on the electric charge shows that the 
strange charge is the dominant order parameter for this transition.
 However, we can expect
that some dependence on the electric charge would arise if charged strange
particles were included, because of the correlation which would then exist
between $n_Q$ and $n_S$.
The middle part of Fig. \ref{Fig1} illustrates the projection of the
phase coexistence domains in the $n_B-n_Q$ plane for  $\mu_S=0$,  
corresponding to
the strangeness-equilibrium condition which is relevant for star matter. 
As discussed above, the coexistence domain being
three-dimensional this representation depends on the value of the third
variable given by $\mu_S$ (or $n_S$). The
well-known isospin dependence of the LG phase transition occurring at $n_S=0$
\cite{Ducoin-06} is apparent.  We have just noticed that the order parameter of
the strangeness phase transition is given by a combination of the strange and
baryonic density. Not surprisingly, the direction of phase separation of this
transition in the $n_B-n_Q$ plane is thus dominated by the baryonic
density. The  order parameter component in the direction of the
electric charge can be understood as due to the correlation between the
different densities. We are facing a transition between a relatively diluted,
non-strange phase to a relatively dense, more strange one. Since $\Lambda$'s
are neutral, the positively charged component of the baryonic density is
relatively less important in the dense phase, which explains the slope of the
separation direction.

\begin{figure}
\begin{center}
\includegraphics[angle=0, width=0.95\columnwidth]{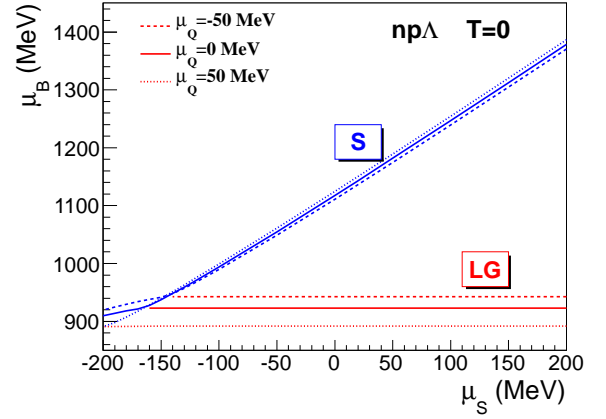}
\end{center}
\caption{(Color online) 
Borders of the phase-coexistence domains corresponding to
the $(n, p, \Lambda)$-mixture at T=0 and $\mu_Q=-50,0,50$ MeV in
$\mu_B-\mu_S$ coordinates.
Red: liquid-gas phase transition of non-strange dilute nuclear matter;
blue: phase transition from non-strange to strange compressed baryonic matter.
} 
\label{fig:phd_npL_mu}
\end{figure}

Fig. \ref{fig:phd_npL_mu} 
offers the complementary image on how the two phase coexistence 
domains look like when plotted with respect to $\mu_S-\mu_B$. 
In this case the condition $\mu_S=0$ is
released and alternative arbitrary
constraints on $\mu_Q=-50, 0, 50$ MeV are imposed.
The absence of $\Lambda$-hyperons at sub-saturation densities makes the
conjugated chemical potential undefined. Mathematically, this means that
any $\mu_{\Lambda} \leq (U_{\Lambda}+m_{\Lambda}c^2)$ is possible. 
This makes $\mu_S$ span a semi-infinite domain lower limited by 
$(\mu_n-U_{\Lambda}-m_{\Lambda}c^2)$.
Reminding that - in $\mu_n-\mu_p$ coordinates - the nuclear matter LG phase 
coexistence is figured by a curve whose extremities are the critical points,
it is easy to understand that, by fixing $\mu_Q$, one fixes both $\mu_n$ and
$\mu_p$. Now, one can straightforwardly identify the semi-infinite horizontal
coexisting lines as the ones corresponding to LG.
The strangeness-driven phase coexistence at fixed  $\mu_Q$ 
appears in  $\mu_S-\mu_B$ as two merged semi-infinite linear segments. 
The merging point corresponds to the state where the 
equilibrium counter-part of the dense phase jumps from
vacuum (low $\mu_S$) to a dilute mixture (high $\mu_S$).

\subsection{Influence of the Coulomb interaction}

We now turn to investigate the influence of Coulomb effects on the 
phase diagram.  
For simplicity, we will consider only electrons and
neglect other charged leptons or mesons.  

\begin{figure}
\begin{center}
\subfloat{
\includegraphics[angle=0, width=0.95\columnwidth]{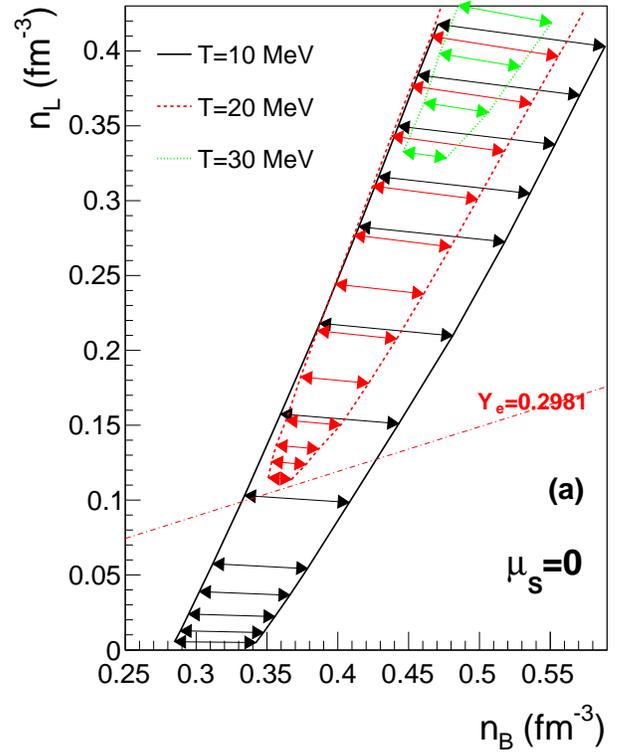}}
\end{center}
\caption{(Color online) 
Borders of the strangeness driven phase transition domain 
corresponding to the neutral net-charge $(n, p, \Lambda, e)$-mixture 
at T=10, 20, 30 MeV and $\mu_S=0$ in $n_B-n_L$ coordinates. 
The dot-dashed line marks, for $T$=20 MeV, a path of constant $Y_e=0.298$.
} 
\label{Fig2}
\end{figure}

\begin{figure}
\begin{center}
\includegraphics[angle=0, width=0.95\columnwidth]{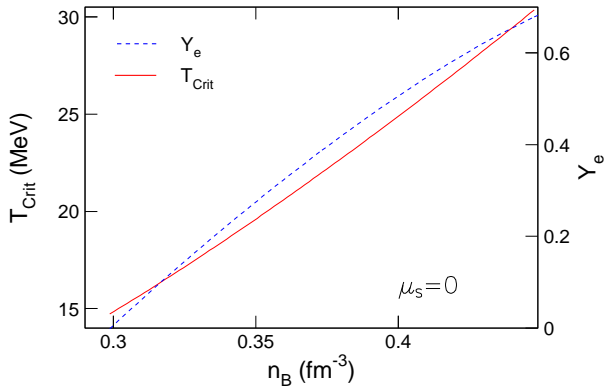}
\end{center}
\caption{(Color online) 
 Electron fraction, $Y_e$, and $n_B$ at the corresponding critical
temperature for $\mu_S=0$.
} 
\label{Fig2bis}
\end{figure}

Electrons are coupled to charged baryons through the electromagnetic
interaction, which can modify the baryonic phase diagram.  
However, the charge neutrality condition
$n_Q=0$ makes the associated chemical potential $\mu_Q$ ill-defined and keeps
the problem three-dimensional~\cite{anomalous}.  Since in homogeneous matter
with the condition $n_Q = 0$ the Coulomb interaction exactly
vanishes \cite{anomalous}, the total mean field pressure can be
written as a sum over independent terms
$P_{B}+P_{L}+P_{\gamma}+P_{\nu}+...$.
We shall still concentrate on the $P_{B}$ contribution, 
as the other terms do
not affect the convexity properties of the thermodynamical potential on which
phase transitions rely.  We have constructed the full phase diagram of the
$(np\Lambda e)$ system in the $(n_B,n_S,n_L)$ space using the hybrid ensemble
\begin{equation}
\bar f_{baryon}(n_B, \mu_S, \mu_L)=
f_{baryon}-\mu_S n_S - \mu_{L} n_{L}.
\end{equation}
In practice, the
charge neutrality condition gives $n_p=n_e$, which allows to infer the
electron chemical potential, $\mu_e$, via
\begin{equation}
n_e=\frac{1}{3 \pi^2} \left( \frac{\mu_e}{\hbar c} \right)^3
\left[1+\mu_e^{-2} \left( \pi^2 T^2 -\frac12 m_e^2 c^4 \right) \right],
\label{eq:mue_ne}
\end{equation}
and to obtain $\mu_L=\mu_p-\mu_n+\mu_e$. Note that $n_e=n_{e^-}-n_{e^+}=n_L(=n_p)$
stands here for the net electron density. 
 If neutrinos were in equilibrium, then $\mu_L$ would correspond to
   the electron neutrino chemical potential. In particular, in a cold neutron
   star in $\beta$-equilibrium we would have $\mu_L = 0$ fixing $n_e$. In core
   collapse events, on the other hand, neutrinos cannot be considered in
   equilibrium in the major part of the system. Here we
 want to study the entire phase diagram, not restricting to $\beta$-equilibrium,
and we therefore leave $\mu_L$ free. Neutrinos, even in equilibrium, would not
change the phase properties, and are thus neglected for the sake of simplicity
in the present discussion.

The lowest panel of Fig. \ref{Fig1} depicts the phase coexistence 
regions of the $(np\Lambda e)$ system at $T=0$ and $\mu_S=0$.  
In agreement with the results of Ref. \cite{Ducoin-07,providencia},
a strong Coulomb quenching of the LG-phase transition is obtained. 
However, the coexistence domain of the
strangeness-driven phase transition is practically un-modified. This can be
easily understood from the fact that the effect of the neutrality condition
$n_Q = 0$ on the two phase transitions is very different. The phase transition
at sub-saturation density has the total baryonic density as order parameter.
At such densities, $n_B$ is strongly correlated to $n_p$ because of the
nuclear symmetry energy which favorizes symmetric $n_n=n_p$ matter. The phase
transition thus implies a discontinuity in $n_p=n_e$, which is strongly
disfavored by the huge electron incompressibility \cite{Ducoin-07}.  At
supersaturation densities the order parameter is given by $n_S$ which is very
loosely correlated to $n_Q$. The phase transition thus does not imply any
strong change in the electron distribution and the presence of electrons thus
does not influence much the phase diagram.

The temperature dependence of the phase diagram along  $\mu_S=0$
is presented in 
Fig. \ref{Fig2}.
We can observe that  the direction of phase separation is almost 
independent of $T$. 
More interesting,  starting from a finite value of $T$, a critical point appears and survives
up to very high temperature.
On 
Fig. \ref{Fig2bis} the critical temperature and 
the electron fraction $Y_e = n_e/n_B$ are shown as a function of
baryon density. These values are typically reached   
within the cooling proto-neutron star, meaning that effects of criticality should be experienced.

\section{Effect of the phase-transition on the neutrino mean free path}
\label{sec:neutrinos}
The cooling of proto-neutron stars is mostly driven by neutrino diffusion 
during the first seconds.
To explore the consequence of criticality for the cooling of proto-neutron star,
we therefore turn to calculate the
mean free path for the neutrino scattering off n, p, and $\Lambda$
particles including the long-range correlations, essential for the
study of criticality, in the linear response approximation.  In
the non-relativistic limit for the baryonic components, the mean free path
at temperature $T$ of a neutrino with initial energy $E_\nu$ is given
by $1/\lambda=1/\lambda^V+1/\lambda^A$ \cite{Iwamoto1982,Navarro1999}
where the contribution of the vector channel is defined as,
\begin{equation}
\frac{1}{\lambda^{V}(E_\nu,T)} = \frac{G_F^2}{16\pi^2} \int (1+\cos\theta)
\mathcal{S}^{V}(q,T) (1-f_\nu(\mathbf{k}_3)) d\mathbf{k}_3,
\label{eq:lambdav}
\end{equation}
and that of the axial channel,
\begin{equation}
\frac{1}{\lambda^{A}(E_\nu,T)} = \frac{G_F^2}{16\pi^2} \int (3-\cos\theta)
\mathcal{S}^{A}(q,T) (1-f_\nu(\mathbf{k}_3)) d\mathbf{k}_3.
\label{eq:lambdaa}
\end{equation}
In Eqs.~(\ref{eq:lambdav})-(\ref{eq:lambdaa}), 
$G_F$ is the Fermi constant, $\theta$ is the 
angle between the initial and final neutrino momentum (=$\mathbf{k_3}$),
$q$ is the transferred energy-momentum, $q=(\omega,\mathbf{q})$, and
$f_\nu$ is the Fermi-Dirac distribution of the outgoing neutrino.
$S^V$ ($S^A$) are the dynamical response function in the vector (axial) channel.
Since this study is focused on the impact of the density fluctuations close 
to the critical point, only the vector channel is considered. 
For densities close to the critical point, spin-density fluctuations are 
however expected to be small \cite{Polls2002,Margueron2003}.

The dynamical response function in the vector channel is defined as
\begin{eqnarray}
\mathcal{S}^{V}(q,T)&=&-\frac{2}{\pi}\frac{1}{1-\exp(-\omega/T)} \times
\nonumber \\
&&\left(\begin{array}{ccc} c_V^n & c_V^p & c_V^\Lambda \end{array} \right)
\Pi^{V}(q,T)
\left(\begin{array}{c} c_V^n\\ c_V^p \\ c_V^\Lambda \end{array} \right),
\end{eqnarray} 
where $\Pi^{V}(q,T)$ is the vector-polarization
matrix for the three species n, p, and $\Lambda$, given by the
Lindhard functions in the case of the mean-field approximation and by
the solution of the Bethe-Salpeter equations in the case of the
mean-field+RPA approximation \cite{Iwamoto1982,Navarro1999,Margueron2003}. 
The vector coupling constants are set to be: 
-1 (n), 0.08 (p), -1 ($\Lambda$)~\cite{Reddy1998}.  
The residual p-h interaction is derived from the potential energy
(\ref{eq:epspot}) and is closely related to the curvature matrix 
without electrons \cite{Ducoin2008}.
 
\begin{figure*}
\begin{center}
\includegraphics[width=.8\textwidth]{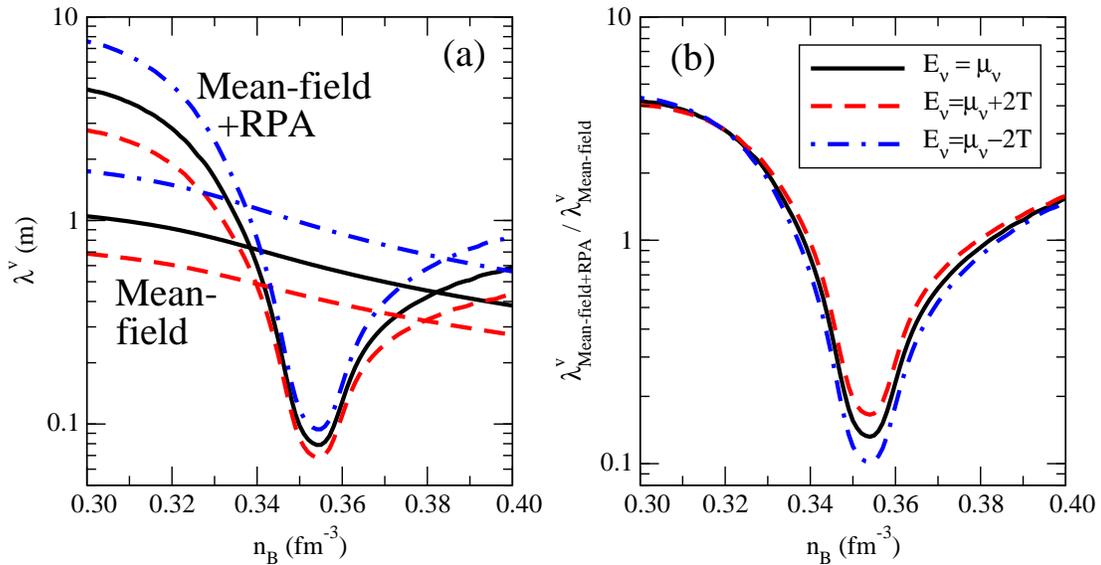}
\end{center}
\caption{(Color online) 
(a) Neutrino mean free path for the scattering off n, p, and $\Lambda$ at
$T=20$ MeV along a constant-$Y_e=0.2981$ trajectory in the phase diagram
for $E_\nu=\mu_\nu$, $\mu_\nu\pm T$ as a function of the
baryonic density $\rho$.
The result of the mean-field approximation is compared to the mean-field+RPA.
(b) The ratio of the mean free path within mean-field+RPA over mean-field
approximation is shown.} 
\label{fig:mfp}
\end{figure*}

The neutrino mean free path along an arbitrary $Y_e=0.2981$ trajectory 
in the phase diagram which passes by the critical point
(see dot-dashed line in Fig. \ref{Fig2}(a)) 
is shown in Fig. \ref{fig:mfp}.
As expected, the RPA correlations strongly reduce the neutrino mean
free path close to the critical point, similar to the critical opalescence 
effect observed for the photon scattering off matter
in critical water.
The ratio of the neutrino mean free path in mean-field+RPA approximation over
that at the mean-field level is shown in panel (b) 
exploring different neutrino energies
around the neutrino chemical potential defined at beta equilibrium.
The effect of the RPA correlations around the critical point is almost 
independent of the neutrino energy in agreement with the interpretation 
as critical opalescence. 

\section{Conclusions}
\label{sec:conclusions}
To conclude, we have studied the phase diagram of a mixture
constituted of interacting neutrons, protons and $\Lambda$-hyperons
under the condition of strangeness-equilibrium, relevant for
supernovae and neutron star physics.  At supra-saturation densities, a
strangeness-driven phase transition can take place, depending on the
assumed strengths of nucleon-$\Lambda$ and $\Lambda$-$\Lambda$
interactions \cite{nL}.  This second transition survives the screening
effect of electrons and persists over a large domain of temperatures
such that it may have an impact on star phenomenology. For a first
study of this equation of state (EoS) within core-collapse supernovae,
see \cite{Peres12}. In addition to the EoS, linear response theory shows that
the neutrino mean-free path dramatically decreases close to the
critical point of this phase transition, which occurs in a
thermodynamic domain accessible to newly-born proto-neutron stars.
 
These results present a first step, and quantitative results might be
somewhat modified in the presence of other strange- and non-strange
baryonic, leptonic or mesonic degrees of freedom.  This work is in
progress and it will make the subject of a forthcoming publication.

\acknowledgments 

This work has been partially funded by the SN2NS project 
ANR-10-BLAN-0503 and it has been supported by
Compstar, a research networking program of the European Science
foundation.
Ad. R. R acknowledges partial support from the Romanian National
Authority for Scientific Research under grants 
PN-II-ID-PCE-2011-3-0092 and PN 09 37 01 05
and kind hospitality from LPC-Caen.

\end{document}